\documentclass[pop,fleqn]{article}
\usepackage{times}
\usepackage{amsmath}
\usepackage{amssymb}
\usepackage{slashed,cite}
\usepackage{w-thm}
\usepackage[dvips]{graphicx}

\newcommand{\Z}{{\mathbb Z}}
\newcommand{\ov}{\overline}
\newcommand{\3}{{\bf 3}}
\newcommand{\2}{{\bf 2}}

\title{Catching the phantom: the MSSM on the $\mathbb{Z}_6$-orientifold}

\author{Tassilo Ott\footnote{Corresponding
     author: e-mail: {\sf tassilo.ott@fys.kuleuven.ac.be}}
     \\Instituut voor Theoretische Fysica,\\ Katholieke Universiteit Leuven,\\
       Celestijnenlaan 200D, B-3001 Leuven,\\ Belgium}

\begin{document}
\maketitle
\begin{abstract}
These lecture notes give a short introduction of the derivation of
the supersymmetric standard model on the
$\mathbb{Z}_6$-orientifold as published in hep-th/0404055.
Untwisted and twisted cycles are constructed and one specific
model is discussed in more detail.
\end{abstract}                   





\section{Introduction}

At the time of the writing of this article, intersecting D6-branes
in type IIA string theory are already a long-studied topic of
perturbative string theory. It started off by the insight that
chiral fermions are possible in these models \cite{Berkooz:1996km}
and now has been proven to be a well-justified complementary
approach to the heterotic string, for a broader introduction see
for instance \cite{Ott:2003yv} and references within.

The goal of all these works is to derive the observed $D=4$ low
energy spectrum of particle physics from string theory. This
requires compactifications with unitary (or orthogonal) gauge
factors with gauge groups of the standard model or a GUT theory.
In fact, only the standard model is well established
experimentally, but extensions like GUT groups [e.g. a flipped
$SU(5)$ or $SO(10)$] seem well motivated such that they also
deserve a discussion from string theory. The same is true for
$N=1$ spacetime supersymmetry, it is not experimentally found, but
very well motivated theoretically and it will be searched for at
the LHC. In case it was found, predictions for certain parameters
of a supersymmetric extension of the standard model from string
theory will be needed. But here one also has to mention the
familiar problem of string theory in making definite statements,
being the large perturbative vacuum degeneracy. In other words, in
many cases there are distinct perturbative string theoretical
models which agree on some established features of the standard
model, but differ in possible extensions. Nevertheless, one hope
is still that some regions of parameter space are at least
excluded on the one hand and that maybe on the other hand some
model features always appear together like for instance three
quark generations with the same rank of the gauge group. It has to
be said that so far, we can only make such statements for a very
specific perturbative theory, like for instance for a certain
$\mathbb{Z}_n$-orientifold in type IIA. In recent times, a new
approach has been proposed by Douglas \cite{Douglas:2003um} which
is now commonly called the landscape. In this picture, the idea is
not anymore to understand why we live in a specific vacuum of
string theory, but instead to use statistics for getting an
overall picture of the landscape of all vacua (which has been
estimated to be at least $10^{100}$ different flux vacua, see
\cite{Douglas:2004kp} and references within).

 The presented paper \cite{Honecker:2004kb} contributes to this
approach in the sense that it gives a complete classification of
models on the $\mathbb{Z}_6$-orientifold, although statistical
tools are actually not needed.

\label{ss:sec1} After having specified a certain perturbative
string theory (here we will use type II plus additional D-branes),
two questions still remain open. One tries to make reasonable
assumptions for these two questions and later tries to justify
them in a bottom-up approach.

The first one asks for the nature of the six-dimensional compact
subspace of the ten spacetime dimensions. Several approaches have
been pursued recently: most universal, general Calabi-Yau spaces
have been treated in \cite{Blumenhagen:2002wn}. In this approach,
there is the problem that generally only the R-R tadpole and the
chiral massless spectrum can be determined by homological data.
The NS-NS tadpoles and the non-chiral massless spectrum cannot be
determined without a CFT calculation which is only in some cases
available. A large subset of such cases are the orientifolded and
orbifolded toroidal models, which will be discussed soon. Another
alternative case where a CFT description is available are the
so-called Gepner or Minimal models which have led to decent
phenomenological models within the last year
\cite{Dijkstra:2004ym, Dijkstra:2004cc}.

The second open question asks which objects are living in
spacetime, meaning in the present context the D-brane and
orientifold content of the theory. Of course, both objects are not
independent of the given spacetime. The orientifold planes, being
non-dynamical objects, actually are completely defined by dividing
out some worldsheet and spacetime groups of the original
spacetime, so this already depends on the answer to the first
question. On the other hand, the D-branes are dynamical objects
and for a complete understanding the backreaction onto spacetime
has to be taken into account. However, it is not necessary for the
calculation of only the tadpole equation and the massless
spectrum. For doing this, D6-brane model building in type IIA
seems to be very attractive as D6-branes  can wrap special
Lagrangian 3-cycles of the compact space. This leads to a very
geometrical picture which will shortly be discussed in the
following section.
\section{D6-brane model building}
The starting point for our considerations is type I theory on a
six-torus $T^6$, whose closed string sector corresponds to type
IIB string theory if the world sheet parity $\Omega$ has been
gauged. Formally, this gauging can be described by the
introduction of orientifold 9-planes. In the language of topology,
this object is a cross-cap (reversing the orientation of the
worldsheet). In this picture, type I theory can be understood as
having a stack of 32 parallel D9-branes whose R-R-charge cancels
the one from the orientifold plane. If one now adds a constant
magnetic F-flux to the system, the requirement to have exactly 32
D9-planes is getting relaxed. In order to perform CFT
calculations, from now on it is assumed that the compact 6-torus
is factorized into three 2-tori, i.e. $T^6=T^2\times T^2\times
T^2$.

\begin{figure}[htb]\label{fig:einf_bsp}
\centering
\includegraphics[width=10cm]{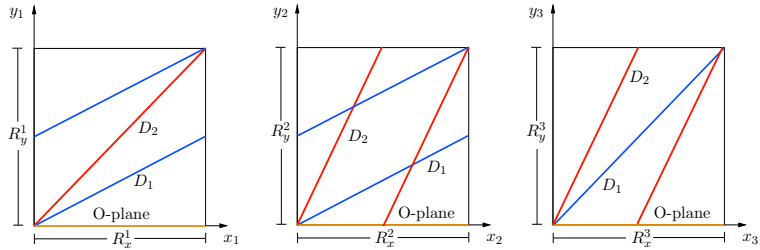}
\caption{A simple example of intersecting D6-branes on the $\Omega
R$-orientifold.}
\end{figure}

\begin{figure}[htb]
\begin{minipage}{.45\textwidth}
\includegraphics[width=\textwidth]{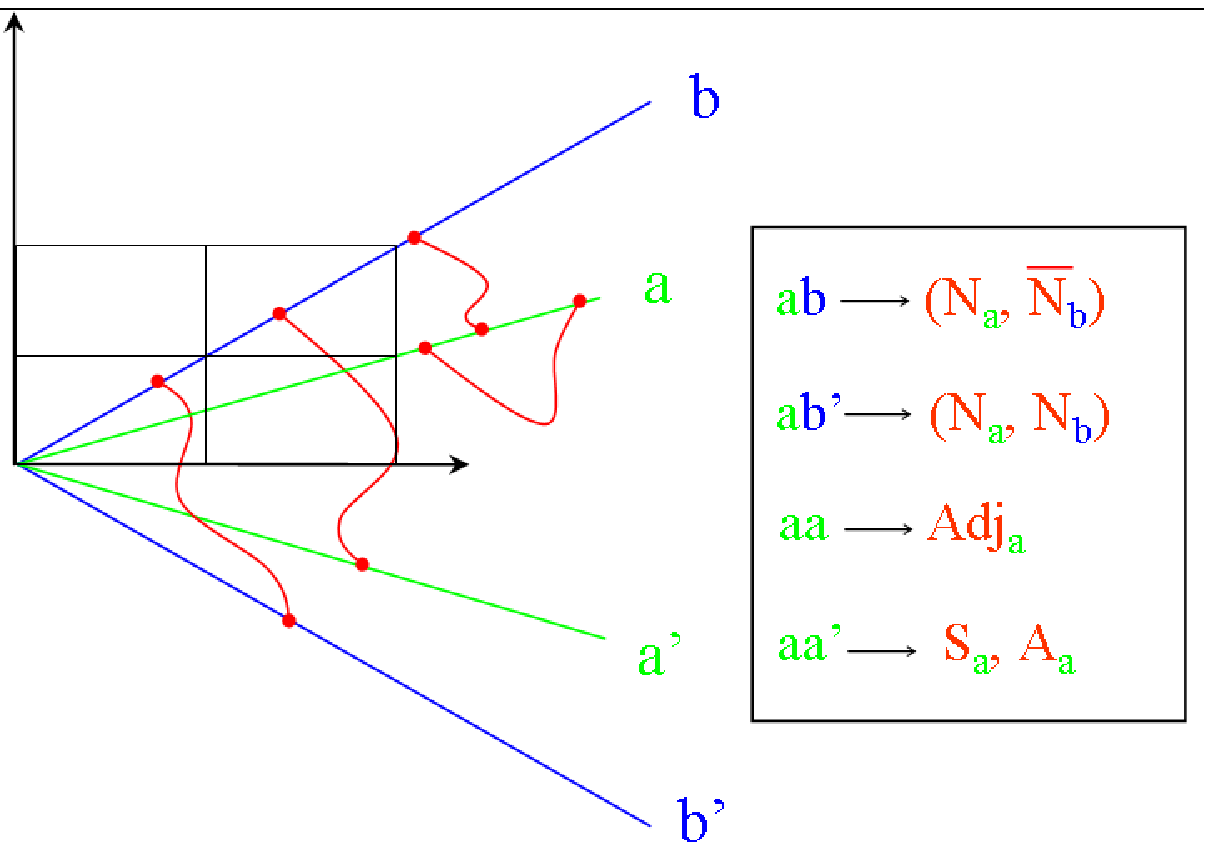}
\caption{Different topological intersections.}
\label{fig:intersections}
\end{minipage}
\hfil
\begin{minipage}{.45\textwidth}
 \caption{The massless chiral open string
spectrum in four dimensions.} \label{tab:massless_chiral_spectrum}
\renewcommand{\arraystretch}{1.5}
\begin{tabular}{ll} \hline
  representation & multiplicity\\
  \hline
  $[{\bf A_a}]_{L}$ & ${1\over 2}\left(\pi'_a\circ \pi_a+\pi_{{\rm O}6} \circ \pi_a\right)$ \\
  $[{\bf S_a}]_{L}$ & ${1\over 2}\left(\pi'_a\circ \pi_a-\pi_{{\rm O}6} \circ \pi_a\right)$ \\
  $[{\bf (\overline{N}_a,N_b)}]_{L}$ & $\pi_a\circ \pi_{b}$ \\
  $[{\bf (N_a, N_b)}]_{L}$ & $\pi'_a\circ \pi_{b}$ \\
  \hline
\end{tabular}
\end{minipage}
\end{figure}

A simpler way to understand this can be obtained by performing
three T-dualities along the y-axes of the three 2-tori. Then the
former F-flux on the D9-branes (which in general is assumed to be
different for some stacks of D9-branes) gets transformed into
D6-branes wrapping complex 1-cycles on every $T^2$, altogether a
complex 3-cycle \cite{Blumenhagen:2000wh}. The angles which every
brane spans with the x-axes of every 2-torus are directly related
to the former F-flux by
\begin{equation}\label{eq:angle_in_terms_of_F}
    \tan \varphi^{I}=F^{I} .
\end{equation}
It can be written in terms of so-called wrapping numbers $n_I$ and
$m_I$ which simply denote the number of times that a certain brane
wraps the two fundamental cycles with radius $R_x^I$ and $R_y^I$
of the $I$th 2-torus, $\tan \varphi^{I}={m_I R_y^I}/{n_I R_x^I}$.
The performed T-dualities furthermore map type IIB into IIA theory
and the world sheet parity $\Omega$ into the combination $\Omega
R$, where $R$ is an anti-holomorphic involution. $R$ is a
spacetime symmetry and can often be defined as e.g. a complex
conjugation on the complex coordinates of the 2-torus, $R:
z_I\rightarrow \bar z_I$. This additional modded out symmetry has
the effect that the orientifold plane does also wrap complex
1-cycles on every 2-torus. D6-branes with different angles in
general do intersect both among themselves and with the
orientifold plane. One obtains a very geometric picture as shown
in figure \ref{fig:einf_bsp}.
In the present case, one has to add a so-called mirror brane for
every D6-brane to the system in order to keep $\Omega R$
invariance. In general, every stack of $N_i$ coinciding D6-branes
at a particular angle supports a $U(N_i)$ gauge factor if it does
not coincide with the O6-plane (if it does also $SO$ and $Sp$
groups are possible). In \cite{Blumenhagen:2002wn} it has been
described that the chiral massless spectrum of a certain model
only depends on the homological data. Actually, the topological
intersection number between different types of stacks of D-branes
with themselves or the O-plane corresponds to the multiplicity of
certain representations, or in other words, the number of particle
generations. This correspondence is stated in table
\ref{tab:massless_chiral_spectrum} and pictured in figure
\ref{fig:intersections}.
Another important insight is the fact that it is possible to
switch on an additional NS-NS 2-form flux $B^I$ in the D9-brane
picture. This translates into tilted 2-tori on the side of
D6-branes, where now generally odd intersections (including three)
are possible \cite{Blumenhagen:2000ea}. As mentioned earlier, the
most important restrictions for model building arise from the R-R
and NS-NS tadpole equations, where the R-R tadpole equation can be
written homologically as
\begin{equation}\label{eq:RRtadpole_top}
\sum_{a=1}^k {N}_a\ \left(\pi_a+\pi'_a\right)-4\pi_{\rm O6}=0\ .
\end{equation}
In the following, only some particular models shall be mentioned
that have been discussed in the immense literature, for more
models see the references in \cite{Ott:2003yv}. These models
generally can be divided into two categories, the ones with $N=0$
and with $N=1$ spacetime supersymmetry.

First the $N=0$ models shall be discussed. The toroidal $\Omega
R$-orientifold has been discussed in \cite{Blumenhagen:2000wh} and
afterwards models with exactly the standard model gauge groups
have been found \cite{Ibanez:2001nd}. But then it was realized
that there are some remaining complex structure moduli in these
constructions which will let the tori collapse
\cite{Blumenhagen:2001te, Blumenhagen:2001mb}. This problem has
been cured in the $\Z_3$-orientifold, where the complex structure
moduli are frozen because of the Hodge number $h_{2,1}=0$, but
still the dilaton instability remains.

In recent times, mainly the $N=1$ models have been discussed,
starting with the $\Z_2\times \Z_2$-orientifold
\cite{Cvetic:2001nr, Cvetic:2001tj}. In these constructions, the
complex structure moduli are unconstrained ($h_{2,1}=3$), but
there is no stability problem as the NS-NS-tadpoles are cancelled.
General problems of these constructions are the presence of exotic
matter (as compared to the MSSM), the need of a hidden brane
sector (stacks of D-branes which do not intersect with the MSSM
ones, but contribute to the tadpole) and the fact that some MSSM
particles might have to be constructed as composite ones. Then
there has been the $\Z_4$-orientifold of
\cite{Blumenhagen:2002gw}. Here, $h_{2,1}=7$ and $h_{1,1}=31$,
implying that there are contributions from $\Z_2$-twisted sectors,
generally leading to fractional D-branes, which first have been
constructed in \cite{Diaconescu:1997br}. In these models, only
mutual intersection numbers $(\pi_a \circ \pi_b,\pi_a' \circ
\pi_b)=\{ (0,0),(\pm 1,0),(0,\pm 1)\}$ are possible, so three
particle generations do not arise. Nevertheless, some
Pati-Salam-models have been obtained which then could lead to a
MSSM-like model after a cascade of non-abelian brane
recombinations. But exotic chiral massless matter also here was
unavoidable. Similar results have been obtained in the $\Z_4\times
\Z_2$-orientifold \cite{Honecker:2003vq}. In the following, the
$\Z_6$-orientifold shall be introduced which leads to more
promising phenomenological models.
\section{The $\Z_6$-orientifold}
\begin{figure}[t]
\begin{center}
\includegraphics[scale=0.6]{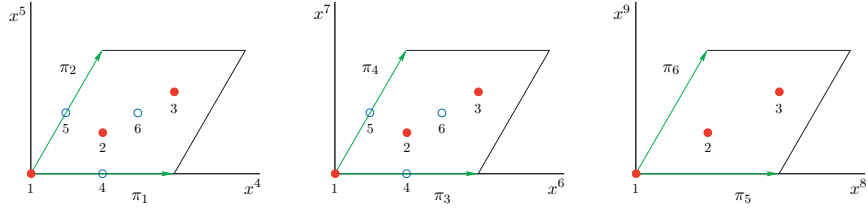}
\end{center}
\caption{Fixed points of the $T^6/\Z_6$ orbifold, depicted for the
{\bf AAA} torus. Full circles denote
  $\theta^2$ fixed points on $T^2_1 \times T^2_2$, empty circles
  additional $\theta^3$ fixed points.}
  \label{Fig:Z6torifixedpoints}
\end{figure}

The discussed model is type IIA on $T^6/ \{\Z_6+\Omega R \Z_6\}$,
where the $\Z_6$ orbifold acts as a rotation $\theta$ on the three
2-tori with angles $\pi/3$, $\pi/3$ and $-2\pi/3$, respectively.
The model is complicated as it involves $\Z_6$-, $\Z_3$- and
$\Z_2$-fixed points on the three 2-tori, as shown in figure
\ref{Fig:Z6torifixedpoints}.
The Hodge numbers are given by $h_{2,1}=5$ and $h_{1,1}=29$,
implying that there are $b_3=2+2h_{2,1}=12$ independent 3-cycles,
out of which two arise directly from the torus geometry, being
denoted by $\rho$. Furthermore, there are ten additional
exceptional 3-cycles $\epsilon$ which are stuck at the
$\Z_2$-fixed points. They have been identified in
\cite{Honecker:2004kb} to be
\begin{equation}
\begin{aligned}\label{Eq:bulkcycles}
\rho_1 &= 2\left( \pi_{1,4,5}+\pi_{1,3,6}+\pi_{2,3,5}-\pi_{1,4,6}-\pi_{2,4,5}-\pi_{2,3,6} \right),\\
\rho_2 &= 2\left(\pi_{1,4,5}+\pi_{1,3,6}+\pi_{2,3,5}-\pi_{1,3,5}-\pi_{2,4,6}\right),\\
\varepsilon_1 &= \left( e_{41}-e_{61}\right) \otimes \pi_5 +\left(
e_{61}-e_{51}\right) \otimes \pi_6, \\ \Tilde{\varepsilon}_1 &=
\left( e_{51}-e_{61}\right) \otimes \pi_5 +\left(
e_{41}-e_{51}\right)
\otimes \pi_6,\\
\varepsilon_2 &= \left( e_{14}-e_{16}\right) \otimes \pi_5 +\left(
e_{16}-e_{15}\right) \otimes \pi_6, \\ \Tilde{\varepsilon}_2 &=
\left( e_{15}-e_{16}\right) \otimes \pi_5 +\left(
e_{14}-e_{15}\right)
\otimes \pi_6,\\
\varepsilon_3 &= \left( e_{44}-e_{66}\right) \otimes \pi_5 +\left(
e_{66}-e_{55}\right) \otimes \pi_6, \\ \Tilde{\varepsilon}_3 &=
\left( e_{55}-e_{66}\right) \otimes \pi_5 +\left(
e_{44}-e_{55}\right)
\otimes \pi_6,\\
\varepsilon_4 &= \left( e_{45}-e_{64}\right) \otimes \pi_5 +\left(
e_{64}-e_{56}\right) \otimes \pi_6, \\ \Tilde{\varepsilon}_4 &=
\left( e_{56}-e_{64}\right) \otimes \pi_5 +\left(
e_{45}-e_{56}\right)
\otimes \pi_6,\\
\varepsilon_5 &= \left( e_{46}-e_{65}\right) \otimes \pi_5 +\left(
e_{65}-e_{54}\right) \otimes \pi_6, \\ \Tilde{\varepsilon}_5 &=
\left( e_{54}-e_{65}\right) \otimes \pi_5 +\left(
e_{46}-e_{54}\right) \otimes \pi_6.
\end{aligned}
\end{equation}
Here, $\pi_{i,j,k}$ just denotes the direct sum of basic
exceptional 1-cycles of the three 2-tori, $\pi_{i,j,k} =\pi_i
\otimes \pi_j \otimes \pi_k$. The symbols $e_{ij}$ denote the
exceptional 2-cycles being stuck at the $i$th $\Z_2$-fixed point
on the first 2-torus and on the $j$th on the second one. The
intersection matrices for both types of cycles are given by
\begin{align}\label{Eq:IntersectionMatrixExcycles}
&I_{\rho}=\left(\begin{array}{cc} 0 & -2
\\ 2 & 0
\end{array}\right),&I_{\varepsilon}=\bigoplus_{j=1}^5 \left(\begin{array}{cc} 0 & 2
\\ -2 & 0
\end{array}\right).
\end{align}
These matrices derive from the fact that on the one hand side the
exceptional 2-cycles $e_{ij}$ always have a self-intersection
number of -2, whereas the intersection numbers of the toroidal
cycles have to be taken to be $\pi_1 \circ \pi_2=+1$, but $\pi_3
\circ \pi_4=\pi_5 \circ \pi_6=-1$. This choice of conventions is
necessary in order to be able to reproduce the homological
computation from the CFT calculation\footnote{We thank to Ralph
Blumenhagen for a valuable discussion on this point.}. From this,
it is possible to construct a basis of fractional cycles,
resembling the topological construction of
\cite{Diaconescu:1997br}. The general construction is described in
\cite{Honecker:2004kb}, only a typical example shall be given
here: a fractional brane can pass through certain $\Z_2$-fixed
points on both the first and second 2-torus, say for instance
fixed point 1 on the first and fixed point 4 on the second torus.
The corresponding exceptional 2-cycle $e_{14}$ together with its
orbifold images $e_{41}$ and $e_{44}$ generate the exceptional
3-cycles $\epsilon_1$, $\epsilon_2$ and $\epsilon_3$. A valid
fractional cycle for this case is given by ${1}/{2}\rho_1 \pm
{1}/{2}\left(\varepsilon_1 \pm \varepsilon_2 \pm \varepsilon_3
\right)$.

As for the toroidal orientifold, there are two possibilities for a
choice of an automorphism of the $\Z_6$-invariant lattice. They
can be understood as two different orientifold projections of the
toroidal 1-cycles, being
\begin{equation}\label{Eq:ORprojectionbulkcycles}
{\bf A}:   \left\{\begin{array}{l}
\pi_{2k-1} \stackrel{R}{\longrightarrow} \pi_{2k-1}, \\
\pi_{2k} \stackrel{R}{\longrightarrow} \pi_{2k-1} - \pi_{2k},
\end{array}\right. \qquad \qquad
{\bf B}:  \pi_{2k-1} \stackrel{R}{\longleftrightarrow}\pi_{2k}.
\end{equation}
This allows for six inequivalent lattices which are all discussed
in \cite{Honecker:2004kb}. Here only the {\bf AAB}-torus will be
mentioned as it gives the most promising results. The orientifold
plane in this case wraps the toroidal cycles $2(\rho_1+\rho_2)$.
The action of the orientifold plane onto the toroidal and
fractional cycles is given by
\begin{align}
    &\rho_1\to \rho_2,& &\rho_2\to \rho_1,& &\varepsilon_1\to -\tilde\varepsilon_1, & &
    \tilde\varepsilon_1\to -\varepsilon_1,& \\
    &\varepsilon_2\to -\tilde\varepsilon_2, & &
    \tilde\varepsilon_2\to -\varepsilon_2,& &\varepsilon_3\to -\tilde\varepsilon_3, & &
    \tilde\varepsilon_3\to -\varepsilon_3,& \nonumber\\
    &\varepsilon_4\to -\tilde\varepsilon_5, & &
    \tilde\varepsilon_4\to -\varepsilon_5,& &\varepsilon_5\to -\tilde\varepsilon_4, & &
    \tilde\varepsilon_5\to -\varepsilon_4.& \nonumber
\end{align}
Another important condition arises if one demands $N=1$
supersymmetry. For the untwisted cycles one only has to demand the
angle criterion. The tree orientated angles which a D-brane
geometrically span (w.r.t. the x-axes of the 2-tori) add up to the
same angle that the orientifold plane is spanning. This just means
that the D-brane has to be calibrated w.r.t. the same holomorphic
3-form as the O-plane in order to be supersymmetric. For the
twisted cycles, the conditions turn out to be slightly more
involved: in simple terms only those $\Z_2$-fixed points are
allowed to contribute which are traversed by the supersymmetric
geometrical part of the brane, for more details see
\cite{Honecker:2004kb}.
\section{The MSSM on the $\Z_6$-orientifold}
In order to find any phenomenologically interesting supersymmetric
models, a computer program has been set up which constructs all
possible fractional cycle configurations for a certain number $n$
of D6-brane stacks. For every SUSY untwisted brane, there are
altogether 16 possibilities to place the brane on the first 2-tori
and switch on Wilson lines and additionally, there are eight
different relative $\Z_2$-eigenvalues. This means that one SUSY
untwisted configuration allows for up to $128^n$ different
supersymmetric fractional brane models. The chiral spectrum for
all these configurations which exactly fulfil the R-R and NS-NS
tadpole conditions has been systematically calculated up to five
stacks. No interesting models with three particle generations and
only bifundamental matter has been found for 2,3 or 4 stacks, but
the case was very different for five stacks. If one demands that
the first three of the five stacks carry a $U(3)$, a $U(2)$ and a
$U(1)$ gauge group, respectively, and that there are exactly three
left handed quark generations in a $(\bar{\mathbf{3}},\mathbf{2})$
representation and that the sum of right handed $U_R$ and $D_R$ in
$(\mathbf{3},1)$ is six, then there remains exactly one chiral
spectrum with just bifundamental matter (although many different
concrete realizations on the {\bf AAB}-torus are possible). This
spectrum is shown in table \ref{Tab:3gen_chiral_AAB321korr}, an
exemplary concrete realization in homology is given in table
\ref{tab:5stackmodel321AABhomologykorr}.
\begin{table}[htb]
\caption{The homology cycles and non-vanishing intersection numbers in the 5 stack
model on the {\bf AAB} torus.}
\label{tab:5stackmodel321AABhomologykorr}\renewcommand{\arraystretch}{1.5}
\begin{tabular}{l|ll} \hline
homology cycles & intersections & \\ \hline
$\Pi_a={1\over 2}\left(
         \rho_1 +\rho_2 +\varepsilon_1 -2\varepsilon_2+\varepsilon_5
         -2\tilde\varepsilon_1+\tilde\varepsilon_2-2\tilde\varepsilon_5\right)$
         &$I_{ab}=0$ &$I_{ab'}=-3$ \\
$\Pi_b={1\over 2}\left(
         \rho_1 +\rho_2 -\varepsilon_1 -2\varepsilon_2-\varepsilon_5+2\tilde\varepsilon_1+\tilde\varepsilon_2+2\tilde\varepsilon_5\right)$
         &$I_{ac}=3$ & $I_{ac'}=3$ \\
$\Pi_c={1\over 2}\left(
         \rho_1 +\rho_2 +3\varepsilon_2 -\varepsilon_4+\varepsilon_5-3\tilde\varepsilon_2-\tilde\varepsilon_4+\tilde\varepsilon_5\right)$
         &$I_{bd}=0$ & $I_{bd'}=3$\\
$\Pi_d={1\over 2}\left(
         \rho_1 +\rho_2 -\varepsilon_1 +2\varepsilon_2-\varepsilon_5+2\tilde\varepsilon_1-\tilde\varepsilon_2+2\tilde\varepsilon_5\right)$
         &$I_{cd}=3$ & $I_{cd'}=-3$\\
$\Pi_e={1\over 2}\left(
         \rho_1 +\rho_2 +3\varepsilon_2 +\varepsilon_4-\varepsilon_5-3\tilde\varepsilon_2+\tilde\varepsilon_4-\tilde\varepsilon_5\right)$
         &$I_{be}=3$ & $I_{be'}=3$ \\ \hline
\end{tabular}
\end{table}
\begin{table}[htb]
\caption{Chiral spectrum of the model with gauge group
$SU(3)_a \times SU(2)_b
  \times U(1)_a \times U(1)_b \times U(1)_c \times U(1)_d \times U(1)_e$.}
\label{Tab:3gen_chiral_AAB321korr}\renewcommand{\arraystretch}{1.0}
\begin{tabular}{llllllllll} \hline
        \multicolumn{10}{c}{\rule[-3mm]{0mm}{8mm}
\text{ chiral spectrum of 5 stack model on the {AAB} torus}} \\
\hline\hline & sector & $SU(3)_a \times SU(2)_b$ & $Q_a$ & $Q_b$
& $Q_c$ & $Q_d$ & $Q_e$ & $Q_{B-L}$& $Q_Y$ \\\hline
$Q_L$ & ab' & 3 $\times$ ($\ov{\3},\2$) & -1 & -1 & 0 & 0 & 0 & ${1}/{3}$& ${1}/{6}$ \\
$U_R$& ac & 3$\times$ ($\3,1$) & 1 & 0 & -1 & 0 & 0 & $-{1}/{3}$& $-{2}/{3}$ \\
$D_R$ & ac' & $3\times$ ($\3,1$) & 1 & 0 & 1 & 0 & 0 & $-{1}/{3}$& ${1}/{3}$  \\
$L$ & bd' & 3 $\times$ ($1,\2$) & 0 & 1 & 0 & 1 & 0 & -1& $-{1}/{2}$ \\
$E_R$ & cd & 3 $\times$ ($1,1$) & 0 & 0 & 1 & -1 & 0 &1 & 1\\
$N_R$ & cd' & 3 $\times$ ($1,1$) & 0 & 0 & -1 & -1 & 0 & 1&0 \\
& be &  3 $\times$ ($1,\2$) & 0 & 1 & 0 & 0 & -1 & 0& 0\\
& be' &  3 $\times$ ($1,\2$) & 0 & 1 & 0 & 0 & 1 & 0 & 0\\
\hline
\end{tabular}
\end{table}
It resembles almost exactly the MSSM, but there are two types of
representations which transform under an additional gauge group
$Q_e$ which still need explanation. The application of the
generalized Green-Schwarz mechanism gives the result that three of
the initial five $U(1)$s are free of triangle anomalies and
massless, being $Q_{B-L}=-{1}/{3}Q_a-Q_d$, $Q_c$ and $Q_e$. The
first one is a $B-L$ symmetry and $Q_c$ is twice the component of
the right-handed weak isospin. $Q_e$ is an additional $U(1)$
symmetry under which only the two additional fields transform, but
none of the standard model particles. The model has a massless
hypercharge which is given by the combination
$Q_Y=-{1}/{6}Q_a+{1}/{2}Q_c-{1}/{2}Q_d$.

The two types of additional particles in the chiral fermion
spectrum can be understood as the supersymmetric standard model
partners of the Higgs fields with a vanishing hypercharge, $H$ and
$\bar H$. This explanation requires an abelian brane recombination
of the two branes $c$ and $e$, triggering the breaking
$U(1)_c\times U(1)_e\rightarrow U(1)_C$ in the effective theory.
It is shown in detail in \cite{Honecker:2004kb} that this
mechanism always works and in the effective theory even can be
understood as a Higgs effect. All these results are indeed very
promising and the phenomenology of the whole class of models with
this chiral spectrum should be further explored. One of the most
burning questions in this context are Yukawa and gauge couplings
\cite{Cvetic:2003ch, Lust:2004cx, Cremades:2004wa}, where it has
to be mentioned that these do depend on the internal geometry and
the full massless spectrum, and therefore are difficult to
calculate. Furthermore, it is likely that they are different for
every concrete realization, maybe a statistical approach could
rather handle this difficulty.

\begin{center}
{\bf Acknowledgements}
\end{center}

I would like to thank G. Honecker and R. Blumenhagen for
interesting discussions and at the same time G. Honecker and J.
Rosseel for proofreading this article. This work is supported in
part by the European Community's Human Potential Programme under
contract MRTN-CT-2004-005104 `Constituents, fundamental forces and
symmetries of the universe'. The work of T.O. is supported in part
by the FWO - Vlaanderen, project G.0235.05 and by the Federal
Office for Scientific, Technical and Cultural Affairs through the
"Interuniversity Attraction Poles Programme -- Belgian Science
Policy" P5/27.

\end{document}